\documentstyle[11pt,aaspp4]{article}
\begin{document}
\slugcomment{AJ, in press (Aug, 2004)}

\title{A Survey of $z>5.7$ Quasars in the Sloan Digital Sky Survey III:
Discovery of Five Additional Quasars\altaffilmark{1}}

\author{Xiaohui Fan\altaffilmark{\ref{Arizona},\ref{KPNO}},
Joseph F. Hennawi\altaffilmark{\ref{KPNO},\ref{Princeton},},
Gordon T. Richards\altaffilmark{\ref{Princeton}},
Michael A. Strauss\altaffilmark{\ref{Princeton}},
Donald P. Schneider\altaffilmark{\ref{PSU}}, 
Jennifer L. Donley\altaffilmark{\ref{Arizona}},
Jason E. Young\altaffilmark{\ref{Arizona}},
James Annis\altaffilmark{\ref{FNAL}},
Huan Lin\altaffilmark{\ref{FNAL}},
Hubert Lampeitl\altaffilmark{\ref{FNAL}}, 
Robert H. Lupton\altaffilmark{\ref{Princeton}}, 
James E. Gunn\altaffilmark{\ref{Princeton}},
Gillan R. Knapp\altaffilmark{\ref{Princeton}}, 
W. N.  Brandt\altaffilmark{\ref{PSU}},
Scott Anderson\altaffilmark{\ref{Washington}},
Neta A. Bahcall\altaffilmark{\ref{Princeton}},
Jon Brinkmann\altaffilmark{\ref{APO}},
Robert J. Brunner\altaffilmark{\ref{uiuc}},
Masataka Fukugita\altaffilmark{\ref{CosmicRay}},
Alexander S. Szalay\altaffilmark{\ref{JHU}},
Gyula P. Szokoly\altaffilmark{\ref{mpe}},
Donald G. York\altaffilmark{\ref{Chicago}}
}

\altaffiltext{1}{Based on observations obtained with the
Sloan Digital Sky Survey,
and with the Apache Point Observatory
3.5-meter telescope,
which is owned and operated by the Astrophysical Research Consortium;
and with the MMT Observatory, a joint facility of the University of
Arizona and the Smithsonian Institution, with the Univeristy of Arizona
2.3-meter Bok Telescope, with the Kitt Peak National Observatory 4-meter Mayall Telescope, with the 6.5-meter Landon Clay Telescope at the
Las Campanas Observatory, a collaboration
between the Observatories of the Carnegie Institution of Washington, University of Arizona, Harvard University, University of Michigan, and Massachusetts Institute of Technology , and with the Hobby-Eberly
Telescope, 
which is a joint project of the University of Texas at Austin,
the Pennsylvania State University, Stanford University,
Ludwig-Maximillians-Universit\"at M\"unchen, and Georg-August-Universit\"at
G\"ottingen.}

\newcounter{address}
\setcounter{address}{2}
\altaffiltext{\theaddress}{Steward Observatory, The University of Arizona,
Tucson, AZ 85721
\label{Arizona}}
\addtocounter{address}{1}
\altaffiltext{\theaddress}{
Visiting Astronomer, Kitt Peak National Observatory, National Optical Astronomy Observatory, which is operated by the Association of Universities for Research in Astronomy, Inc. (AURA) under cooperative agreement with the National Science Foundation. 
\label{KPNO}}
\addtocounter{address}{1}
\altaffiltext{\theaddress}{Princeton University Observatory, Princeton,
NJ 08544
\label{Princeton}}
\addtocounter{address}{1}
\altaffiltext{\theaddress}{Department of Astronomy and Astrophysics,
The Pennsylvania State University,
University Park, PA 16802
\label{PSU}}
\addtocounter{address}{1}
\altaffiltext{\theaddress}{Fermi National Accelerator Laboratory, P. O. Box 500, Batavia, IL 60510
\label{FNAL}}
\altaffiltext{\theaddress}{University of Washington, Department of Astronomy,
Box 351580, Seattle, WA 98195
\label{Washington}}
\addtocounter{address}{1}
\altaffiltext{\theaddress}{Apache Point Observatory, P. O. Box 59,
Sunspot, NM 88349-0059
\label{APO}}
\altaffiltext{\theaddress}{Institute for Cosmic Ray Research, University of
Tokyo, Midori, Tanashi, Tokyo 188-8502, Japan
\label{CosmicRay}}
\addtocounter{address}{1}
\altaffiltext{\theaddress}{Dept. of Astronomy \& NCSA, University of Illinois,
1002 W. Green Street, Urbana, IL 61801
\label{uiuc}}
\addtocounter{address}{1}
\altaffiltext{\theaddress}{
Department of Physics and Astronomy, The Johns Hopkins University,
Baltimore, MD 21218, USA
\label{JHU}}
\addtocounter{address}{1}
\altaffiltext{\theaddress}{Max-Planck-Institut f\"ur extraterrestrische Physik, Postfach 1312, 85741 Garching, Germany
\label{mpe}}
\addtocounter{address}{1}
\altaffiltext{\theaddress}{University of Chicago, Astronomy \& Astrophysics
Center, 5640 S. Ellis Ave., Chicago, IL 60637
\label{Chicago}}

\begin{abstract}
We present the discovery of five new quasars
at $z>5.7$, selected from the multicolor imaging data
of the Sloan Digital Sky Survey (SDSS).
Three of them, 
at redshifts 5.93, 6.07, and 6.22, 
were selected from  $\sim$ 1700
deg$^2$ of new SDSS Main Survey imaging in the Northern Galactic Cap.
An additional quasar, 
at redshift 5.85, 
was discovered by coadding the data obtained in the Fall Equatorial Stripe
in the SDSS Southern Survey Region.
The fifth object, 
at redshift 5.80, is selected from 
a non-standard SDSS scan in the Southern Galactic Cap outside the Main Survey
area.
The spectrum of SDSS J162331.81+311200.5 ($z=6.22$) shows a complete
Gunn-Peterson trough at $z_{abs} > 5.95$, similar to the troughs
detected in other three $z\gtrsim 6.2$ quasars known.
We present a composite spectrum of the $z>5.7$ quasars discovered
in the SDSS to date.
The average emission line and continuum properties of $z\sim 6$ quasars exhibit
no significant evolution compared to those at low redshift.
Using a complete sample of nine $z>5.7$ quasars, we find that the density of quasars with $M_{1450} < -26.7$
at $z\sim 6$ is $(6\pm2) \times 10^{-10}$ Mpc$^{-3}$ 
($\rm H_0 = 65\ km\ s^{-1} Mpc^{-1}$, $\Omega = 0.35$ and $\Lambda = 0.65$), 
consistent with our previous estimates.
The luminosity distribution of the sample is fit with a power law luminosity function
$\Psi(L) \propto L^{-3.2\pm0.7}$, somewhat steeper than but consistent with our
previous estimates.
\end{abstract}

\section{Introduction}
This paper is the third in a series presenting $i$-dropout
$z\gtrsim 5.7$ quasars selected from the multicolor imaging data
of the Sloan Digital Sky Survey (SDSS; \cite{York00}, Stoughton et al. 2002, 
Abazajian et al. 2003, 2004).
In \cite{z58} and in the first two papers of this series
(\cite{PaperI}, Paper I, \cite{PaperII}, Paper II),
we presented the discovery of seven luminous quasars
at $z=5.74 - 6.42$, selected from $\sim 2900$ deg$^2$ of 
SDSS imaging in the Northern Galactic Cap.
In this paper, we describe the discovery of 
five quasars at $z=5.80$, 5.85, 5.93, 6.07 and 6.22, respectively.
The scientific objectives, photometric data reduction, target
selection and follow-up observation procedures are described
in detail in Paper I.
Three of the new quasars were selected using the same color selection
procedures described in Paper II, as outlined
briefly in \S 2.1. 
One object
was selected using the co-added catalog from multi-epoch imaging
of the Fall Equatorial Stripe in the SDSS Southern Survey region.
The co-addition procedures are described in \S 2.2.
The final object was selected in a non-standard SDSS scan in the Southern
Galactic Cap outside the main survey area.
We present the spectroscopic follow-up observations of the 
$i$-dropout candidates and the photometric and spectroscopic properties
of the newly-discovered quasars in \S 3.
Combining the quasars in this paper and those presented
in Fan et al. (2002) and in Papers I and II, we construct
the composite spectrum of $z\sim 6$ quasars, and compare it with
the average quasar spectrum at low redshift (\S 4). Finally, we update
our estimate of the evolution of the high-redshift quasar luminosity 
function in \S 5.

Following the previous papers in this series, we use two cosmologies to
present our results:
(1) $\rm H_0 = 50\ km\ s^{-1}\ Mpc^{-1}$, $\Omega_{\Lambda} = 0$ and
$\Omega_{M} = 1$ ($\Omega$-model); (2)
$\rm H_0 = 65\ km\ s^{-1}\ Mpc^{-1}$, $\Omega_{\Lambda} = 0.65$ and
$\Omega_{M} = 0.35$ ($\Lambda$-model).
Coincidentally, the luminosity and number densities measured under our $\Lambda$-model
are within 1\% of those in the WMAP (\cite{Spergel03})
cosmology with $\rm H_0 = 71\ km\ s^{-1}\ Mpc^{-1}$, $\Omega_{\Lambda} = 0.73$ and
$\Omega_{M} = 0.27$. 

\section{Candidate Selection and Identification}
The Sloan Digital Sky Survey is using
a dedicated 2.5m telescope and a large format CCD camera (\cite{Gunnetal})
at the Apache Point Observatory in New Mexico
to obtain images in five broad bands ($u$, $g$, $r$, $i$ and $z$,
centered at 3551, 4686, 6166, 7480 and 8932 \AA, respectively; \cite{F96})
of high Galactic latitude sky in the Northern Galactic Cap.
About $6000$ deg$^2$ of sky have been imaged at the
time of this writing (Mar 2004).
The imaging data are processed by the astrometric pipeline
(\cite{Astrom}) and photometric pipeline (\cite{Photo}),
and are photometrically calibrated to a standard star network
(\cite{Smith02}, see also \cite{Hogg01}).
In addition to the Main Survey, the SDSS also obtains multi-epoch 
imaging in $\sim 270$ deg$^2$ along the Celestial Equator in the
Southern Galactic Cap (the SDSS Southern Survey, York et al. 2000).
At the time of this writing, the Southern Survey region has been imaged
between 5 and 15 times,
depending on the RA of the field. The multi-epoch data are used to study variable objects
and are co-added to reach fainter limiting magnitudes. 
We will use imaging from both the Main Survey and the Southern Survey 
to search for quasar candidates.

\subsection{Selection in the Main Survey Area}

The SDSS quasar target selection pipeline (Richards et al. 2002)
selects only quasar candidates at $z\lesssim 5.5$.
The highest-redshift object discovered in the SDSS spectroscopic
survey is at $z=5.4$ (Anderson et al. 2001).
At higher redshift, quasars become $i$-dropout objects; their
selections require additional follow-up observations.
Papers I and II present the results from a survey of $i$-dropout
candidates selected from $\sim 2900$ deg$^2$ of SDSS Main Survey imaging carried
out in the Springs of 2000, 2001 and 2002.
In Spring  2003, we searched for $i$-dropout quasar
candidates in 37 new SDSS imaging runs.
These imaging data were taken between 10 October 2002 (Run 3358)
and 29 April 2003 (Run 3919).
We used the same criterion to decide which photometric runs
to include in the $i$-dropout survey as in Papers I and II:
the $z$ band image quality, measured by
the psfWidth parameter ($=1.06$ FWHM for
a Gaussian profile) in the fourth column of the SDSS camera,
should be  better than 1.8$''$.
The median seeing in the $i$ and $z$ bands is $\sim 1.4''$
for the entire survey area (\cite{DR1}).
There is overlap between adjacent SDSS strips and stripes (York et
al. 2000), meaning that these 37 new runs overlap somewhat with the 
area covered in Papers I and II. 
Taking these overlaps into account, we find that the total
{\em new} area of the sky covered by these runs is
1708 deg$^2$, bringing the combined sky coverage of Papers I, II and
this paper to 4578 deg$^2$.

%
%
%

We applied the same color selection criteria as in Paper II (see Figure 1 and
2 in Paper II) to the new SDSS imaging data to selection $z>5.7$ quasar
candidates.
A total of 80 $i$-dropout candidates were selected in the main survey area.
The photometric and spectroscopic follow-up observations
were carried out over a number of nights between December 2002 and June 2003.
Independent $z$ photometry was carried out using the
Seaver Prototype Imaging camera (SPICAM) in the SDSS $z$ filter on
the ARC 3.5m telescope at the Apache Point Observatory.
$J$ photometry was carried out using the $256 \times 256$ NICMOS imager
on Steward Observatory's 2.3m Bok Telescope at Kitt Peak, and using
the GRIM II instrument 
(the near infrared GRIsm spectrometer and IMager), also on the ARC 3.5m.
The  spectroscopic follow-up observations were obtained
using the Red-Channel Spectrograph on the MMT 6.5m telescope on Mt. Hopkins,
the Double Imaging Spectrograph (DIS) on the ARC 3.5m,
the Multi-Aperture Red Spectrograph (MARS) on the 4-m telescope on Kitt Peak,
and the Low Resolution Spectrograph (LRS, \cite{LRS}) on
the Hobby-Eberly Telescope.

\subsection{Selection in the SDSS Southern Survey Region}
The SDSS Main Survey imaging consists of single epoch observations with
exposure times of 54.1 seconds. 
For the average seeing conditions (FWHM $\sim 1.4''$), 
the 5-$\sigma$ limiting magnitudes in the $i$ and $z$ bands are 22.5 and 20.5,
respectively (Stoughton et al. 2002, \cite{DR1}).
At this depth, the SDSS Main Survey only allows selection of
$i$-dropout candidates at $z<20.2$, and therefore is only sensitive 
to the most luminous quasars at $z\sim 6$ ($M_B \lesssim -27$). 
The multi-epoch imaging obtained in the SDSS Southern Survey area 
will eventually provide photometry $\sim 1.5$ magnitudes deeper than
the main survey over $\sim$ 270 deg$^2$, enabling selection of
much fainter quasar candidates.
By the end of 2003, the SDSS Southern Survey Region had been scanned
5 -- 15 times, resulting in a dataset that goes more than
one magnitude deeper than the Main Survey.

In Fall 2002, we used a preliminary co-added catalog of
$\sim 5$ epoch imaging to select $z\sim 6$ quasar candidates.
The co-added catalog was generated by matching sources detected in
more than one SDSS run within a radius of $1''$.
The average flux is calculated from the SDSS asinh magnitudes (\cite{Luptitude}),
weighted by the inverse of the flux variances.
Note that when co-adding at the catalog level, an object is
required to be detected in the individual
runs. 
Therefore, while the co-addition improves the S/N of faint sources, allowing
us to select candidates all the way down to the 6-$\sigma$ detection limit,
it does not allow
measurement of objects fainter than the detection limit of single
SDSS exposures.
Therefore, we can only select candidates at $z<20.5 - 20.7$.
We searched for candidates from $\sim 100$ deg$^2$ of co-added catalogs.
The color selection criteria in the Southern Survey region
are the same as those used in the 
Main survey area. 
A number of faint $i$-dropout candidates were observed using Magellan
II Clay Telescope 
and the Boller and Chivens (B\&C) spectrograph on 16 Oct 2002.
One quasar, SDSS J000552.34--000655.8\footnote{The naming convention for SDSS
sources is SDSS JHHMMSS.SS$\pm$DDMMSS.S, and the positions are expressed in
J2000.0 coordinates. The astrometry is accurate to better than $0.1''$
in each coordinate.} 
(hereafter SDSS J0005--0006, $z_{AB} = 20.5$) was discovered at
a redshift of 5.85.

\subsection{Selection in the Constant-Longitude Scans}
The SDSS has obtained some imaging data outside the survey
boundary  (York et al. 2000; Finkbeiner et al. 2004), including a few
strips in the Southern Galactic Cap that extend to low Galactic
latitude. 
We have also selected $i$-dropout candidates in these runs.
One new quasar, SDSS J000239.39+255034.8 ($z=5.80$), was selected from Run 4152,
a constant-longitude scan (at $l = 110^\circ$),
observed on 29 Sep 2003.  The sample in these scans which go outside
the main survey area is not yet complete, and so we do not include
this object in our complete sample. 


\section{Discovery of Five New Quasars at $z>5.7$}

Among the 80 $i$-dropout candidates
candidates in the Main Survey area,
15 are false $z$ band only detections which are most likely cosmic rays;
55 are M or L dwarfs (mostly classified
photometrically based on their red $z^* - J$ colors);
and 7 are likely T dwarfs.
Several objects still lack
proper infrared spectroscopy, so the T dwarf classification
is still preliminary. 
Three of the candidates are identified as quasars at $z>5.7$:
SDSS J141111.29+121737.4 ($z=5.93$, SDSS J0002+2550 hereafter),
SDSS J160254.18+422822.9 ($z=6.07$, SDSS J1602+4228),
and SDSS J162331.81+311200.5 ($z=6.22$, SDSS J1623+3112).
The discovery spectra of the first two objects were obtained
using DIS on the ARC 3.5m in April 2003, and the discovery spectrum of
the last object was obtained in June 2003 at MMT using the
Red Channel spectrograph.
We have subsequently obtained longer exposures of these three
quasars using the Red Channel on MMT and MARS on the Kitt Peak 4-meter,
with total exposure times of 2 -- 4 hours each. 
These spectra are shown in Figure 2.
As discussed above, SDSS J000552.34--000655.8 ($z=5.85$) was discovered in the deep imaging of 
the SDSS Southern Survey region. 
Figure 2 shows the discovery spectrum, a 40-min exposure obtained with
the Clay telescope and B\&C spectrograph in Oct 2002.
The discovery spectrum of SDSS J000239.39+255034.8 ($z=5.80$, \S2.3) 
was obtained at the MMT using the Red Channel Spectrograph
in Nov 2003. This one hour exposure is shown in Figure 2. 
 
The finding charts  of the five new quasars
are presented in Figure 1.
The spectra are flux-calibrated to match the observed $z$ band
photometry.
Table 1 presents the photometric properties of the new quasars,
and Table 2 presents the measurements of their continuum properties.
Following Papers I and II, the quantity
$AB_{1280}$ is defined as the AB magnitude of the continuum
at rest-frame 1280\AA, after correcting for interstellar
extinction using the map of \cite{Schlegel98}.
We extrapolate the continuum to rest-frame 1450\AA,
assuming a continuum shape $f_\nu \propto \nu^{-0.5}$, to
calculate $AB_{1450}$.
None of the five quasars is detected in the FIRST (\cite{FIRST}) 
or NVSS (\cite{NVSS}) radio surveys.
The discovery of these five new quasars brings the total of $z > 5.7$
quasars known to twelve, all selected from SDSS imaging.

\subsection{Notes on Individual Objects}

\noindent
{\bf SDSS J000239.39+255034.8 ($z=5.80$).}
This object is selected in a non-standard SDSS scan 
in the Southern Galactic Cap.
It is outside the main survey region, and the spectroscopic
follow-up of this region for $i$-dropout candidates 
is not yet complete. We therefore
do not include it in the luminosity function calculations
below. 

SDSS J0002+2550 has $i=21.47$ and $z=18.99$, but is undetected in
2MASS (\cite{2MASS}), indicating that $J>16.5$ and $z-J<2.5$. If it were an L dwarf,
it would have a $z-J$ color of 2.5 or redder, so we targeted this
object as a high-redshift quasar candidate. 
SDSS J1044--0125 ($z=5.74$, Fan et al. 2000) was selected in a
similar manner.

At $z_{AB}\sim 19$ and $M_{1450} = -27.55$ ($\Lambda$-model), SDSS J0002+2550
is very luminous. It is the second brightest quasar known
at $z>5.7$ so far and provides an excellent target for high
resolution spectroscopic follow-up observations.
The redshift of SDSS J0002+2550 is determined by the locations
of the peak of the Ly$\alpha$ emission line and of
the OI 1300\AA\ emission line, and is accurate to 0.02.
The quasar has a Ly$\alpha$+NV emission line rest-frame equivalent
width of $\sim 60$\AA, comparable to that of most high-redshift
quasars (Fan et al. 2001b).  

\noindent
{\bf SDSS J000552.34--000655.8 ($z=5.85$).}
This object is selected in the SDSS Southern Survey region.
At $z_{AB} = 20.54\pm 0.10$, it is the faintest $z>5.7$ quasar
in our sample.
Although the discovery spectrum (Figure 2) has low S/N,
the strong Ly$\alpha$
and NV emission lines and the strong Lyman break are clearly visible.
The redshift is determined by the locations of
the Ly$\alpha$ and NV emission line peaks, and is accurate to 
0.02. The emission lines in this quasar appear to be quite narrow:
the Ly$\alpha$ and NV emission are clearly separated. We estimate a 
FWHM of about 1500 -- 2500 km\ s$^{-1}$ for
individual lines, with large uncertainties due to the low S/N.

\noindent
{\bf SDSS J141111.29+121737.4 ($z=5.93$).}
The redshift  is determined by the locations of
the OI 1300\AA\ and NV 1240\AA\ lines and is accurate to
0.02. The object has a moderately
strong Ly$\alpha$+NV line (rest-frame equivalent width of $\sim 100$\AA).
This object is detected in two overlapping SDSS runs with consistent photometry
(Table 1).

\noindent
{\bf SDSS J160254.18+422822.9 ($z=6.07$).}
The redshift of this object is  determined by the peaks of
the Ly$\alpha$ and NV 1240\AA\ lines and is accurate to
0.02. A weak OI 1300\AA\ line is also detected in the spectrum.
The quasar spectrum shows a number of dark patches in the Ly$\alpha$
and Ly$\beta$ forest, although the S/N of the current spectrum
is not sufficient to determine whether there is a short Gunn-Peterson
trough in the spectrum.

Due to its proximity on the sky to the high-redshift cluster candidate
CL1603, SDSS~J1602+4228 serendipitously lies in a 28~ks pointed observation 
(sequence rp800239) made with the {\it ROSAT\/} Position Sensitive 
Proportional Counter (PSPC). We do not find any significant X-ray detection 
of SDSS~J1602+4228 in these data. The $3\sigma$ upper limit on its 
observed-frame, Galactic absorption-corrected, 0.5--2~keV flux is
$5.9\times 10^{-14}$~erg~cm$^{-2}$~s$^{-1}$, adopting a power-law model with 
a photon index of $\Gamma=2$ and the Galactic column density of 
$N_{\rm H}=1.3\times 10^{20}$~cm$^{-2}$. Given the $AB_{1450}$ magnitude 
of SDSS~J1602+4228, comparison with Figure~2 of Vignali et~al. (2003) shows 
that the X-ray upper limit is consistent with X-ray observations of other 
$z>4$ quasars. The slope of a nominal power law between rest-frame 2500~\AA\ 
and 2~keV is constrained to be $\alpha_{\rm ox}>1.2$. 

\noindent
{\bf SDSS J162331.81+311200.5 ($z=6.22$).}
This is the highest redshift quasar presented in this paper,
and is the third highest redshift quasar yet known.
\footnote{SDSS J104845.05+463718.3 was reported to have a redshift of 6.23
in Fan et al. (2003). Subsequent observations show that it is
a Broad Absorption Line (BAL) quasar (Maiolino et al. 2004, Fan et al. in preparation) and
the original redshift determination is biased; the best redshift estimate
is $z=6.18$ based on new observations.}
This quasar has two striking features. First, it has an 
extremely strong Ly$\alpha$ emission line. The total equivalent
width of Ly$\alpha$+NV is $\gtrsim 150$\AA\ in the rest-frame,
{\em without} taking into account the strong absorption due
to the Ly$\alpha$ forest on the blue side of the Ly$\alpha$ emission.
For comparison, Fan et al. (2001b) measured the mean and standard deviation of the
rest-frame Ly$\alpha$+NV equivalent width of $69.3 \pm 18.0$\AA, based
on a sample of $\sim 40$ quasars at $z\sim 4$ selected from the SDSS (see also  
Schneider, Schmidt \& Gunn 1991 and \S 4).
The line strength of this quasar is more than a factor of two larger
than the average at high redshift. 
It is one of the strongest-lined quasars at $z>4$ yet known, 
and the strongest-lined quasar at $z>5$.

Second, SDSS J1623+3112 has a complete Gunn-Peterson (1965) trough. 
Following Becker et al. (2001), Fan et al. (2002) and White et al.
(2003), we define the transmitted flux ratio as:
\begin{equation} 
{\cal T}(z_{abs}) \equiv \left\langle f_\nu^{obs}/f_\nu^{con} \right\rangle,
\hspace{1cm} (1+z_{abs}-0.1)\times 1216 \hbox{\AA} < \lambda <
(1+z_{abs}+0.1) \times 1216 \hbox{\AA},
\end{equation}
where $f_\nu^{con}$ is the continuum level extrapolated from the
red side of Ly$\alpha$ emission.
Using the spectrum presented in Figure 2, we find that at $z = 6.05$,
the transmitted flux ratio ${\cal T}(z_{abs} = 5.95 - 6.15) = 0.004 \pm 0.008$, consistent
with zero flux in the Gunn-Peterson trough region. 
We also detect a complete Ly$\beta$ Gunn-Peterson trough in this quasar.
This is the fourth quasar with a complete Gunn-Peterson trough,
after SDSS J1306+0524 ($z=6.28$, Becker et al. 2002), 
SDSS J1148+5251 ($z=6.42$, White et al. 2003) and SDSS J1048+4637 
($z=6.18$, Fan et al. 2003). 

In {\em all quasars at
$z>6.1$, complete Gunn-Peterson troughs are detected}, starting 
from $z_{abs}=5.95 \pm 0.1$,
and extending to the highest redshift not affected by the quasar proximity effect.
The detection of a complete Gunn-Peterson trough in SDSS J1623+3112
further confirms the rapid transition of the ionization state 
of the IGM at $z\sim 6$
(e.g. Becker et al. 2001, Djorgovski et al. 2001, Fan et al. 2002,
White et al. 2003; see also 
Songaila 2004 for a different interpretation).
The S/N of the spectrum of SDSS J1623+3112 is considerably lower than
that of the other three quasars, therefore the optical depth limit we
are able to place is not yet very strong.
In a subsequent paper, we will present a detailed analysis of the constraints on 
the evolution of IGM properties using the five new quasars presented in this paper.

\section{Quasar Composite Spectrum at $z\sim 6$}
The spectral energy distributions of luminous quasars show little
evolution out to high redshift.
There is  growing evidence from emission line ratio measurements
that quasar broad emission line regions have roughly solar or even
higher metallicities at $z>4$ (e.g., \cite{HF93}, Dietrich et al. 2003a), similar to that
in low redshift quasars.
Dietrich et al. (2003b) found the FeII/MgII ratio 
to have roughly the same value in a sample of $z\sim 5$ quasars as at lower redshift,
suggesting that the metallicity of quasar emission line region remains high to
even earlier epochs.

The sample of twelve quasars at $z>5.7$ from the SDSS provides the first opportunity
to study the evolution of quasar spectral properties at $z\sim 6$,
less than 1 Gyr after the Big Bang and only
700 million years from the first star formation in the Universe (Kogut et al. 2003, 
Spergel et al. 2003).
Optical and infrared spectroscopy of some $z\sim 6$ quasars 
already indicates {\em a lack of evolution}
in the spectral properties of these luminous quasars:
Pentericci et al. (2002) show that the CIV/NV ratio in two $z\sim 6$ quasars
are indicative of supersolar metallicity in these systems.
Freudling et al. (2003)  and Barth et al. (2003) 
detected strong FeII emission in the spectra of four $z\sim 6$ SDSS quasars.
In addition, the optical-to-X-ray flux ratios and X-ray continuum
shapes show at most mild evolution from low redshift (e.g. Brandt et
al. 2002, Vignali et al. 2003). 
These results, if confirmed with a larger sample, suggest that the accretion disk
and photoionization structure
of quasars reached maturity very early on and are probably insensitive
to the host galaxy environment.

Figure 3 shows the composite of eleven of our twelve $z>5.7$ quasar
spectra.
We omit SDSS J0005-0006 due to its low S/N.
To produce the composite, we simply redshift all the spectra
to zero, scale the continuum level 
of each quasar based on its $m_{1450}$ magnitude, and 
average all the scaled spectra with equal weighting.
The composite covers rest-frame wavelengths from 1100\AA\ to 1450\AA.
Also plotted in Figure 3 is the low-redshift 
SDSS quasar composite of Vanden Berk et al. (2001).
The effective redshift of the low-redshift composite in this redshift range
is $z\sim 2$.
Blueward of Ly$\alpha$ emission, the strong IGM absorption at $z\sim 6$
almost completely removes the quasar flux.
Redward of Ly$\alpha$ emission,
there is no detectable difference in the {\em intrinsic UV spectral properties}.
The continuum shape is consistent with the power law $f_{\nu} \propto 
\nu^{-0.4}$ measured by  Vanden Berk et al. (2001).
Clearly, spectral coverage in the near-IR is needed to put stronger
constraints on the continuum shape.
Pentericci et al. (2003) use IR photometry of a sample of quasars at
$z = 3.5 - 6$ to measure a continuum shape
of $f_{\nu} \propto \nu^{-0.5}$, independent of redshift. 
The strengths of the emission lines at $z\sim 6$, including NV 1240\AA, OI 1300\AA, CII 1335\AA,
and SiIV+OIV 1400\AA, are also comparable to those at low redshift.
This composite does not go red enough to cover the CIV 1549\AA\ line,
so we cannot test whether the 
Baldwin (1977) effect exists at these redshift.
The strength of the red wing of the Ly$\alpha$ emission line shows no
evolution from that at low redshift.
The weaker Ly$\alpha$ emission in the blue wing is 
due to the strong IGM absorption.
The average FWHM of emission lines is $\sim 6000$ km s$^{-1}$, also consistent
with the low-redshift average.

\section{Luminosity Function of $z\sim 6$ Quasars}

In Papers I and II, we estimated the comoving density of quasars at $z\sim 6$
using a sample of six quasars, covering a total area of 2870 deg$^2$.
We repeat the calculation here, including the additional three
quasars in the complete sample;
as explained above, SDSS 0002+2550 and SDSS J0005-0006 were not selected
as part of the flux-limited complete sample and will not be included
in the quasar luminosity function calculation.
These nine quasars form a complete sample
satisfying  the selection criteria in Eq. (1) over
a total area of 4578 deg$^2$.
Following Papers I and II, 
we calculate the selection function of
$z\sim 6$ quasars using a Monte-Carlo simulation of quasar
colors, taking into account the distribution of quasar emission line
and continuum properties, Ly$\alpha$ absorption, the SDSS
photometric errors and Galactic extinction.
The selection function as a function of redshift $z$ and
absolute magnitude $M_{1450}$ is illustrated in Figure 7 of Paper II
for the $\Lambda$-model.
The total
spatial density of quasars at $z\sim 6$ is derived using the $1/V_{a}$ method.
We find  that at
the average redshift of $\langle z \rangle = 6.07$,
$\rho (M_{1450} < -26.4) = (10.5 \pm 3.9) \times 10^{-10}$ Mpc$^{-3}$
for the $\Omega=1$ model,
and $\rho (M_{1450} < -26.7) = (6.4  \pm 2.4) \times 10^{-10}$ Mpc$^{-3}$
for the $\Lambda$-model.
The results, which are  consistent with those in Paper II with smaller
error bars, are plotted in Figure 4, together with
the measurements at lower redshifts 
from the 2dF survey (\cite{2dF}, Croom et al. 2004) at $z<2.2$,
from \cite{SSG} at $2.7 < z < 4.8$ and from Fan et al. (2001a) at
$3.6 < z < 5.0$.
The comoving density of luminous quasars at $z\sim 6$ is 30 times
smaller than that at $z\sim 3$.

Following Paper II, we derive the
bright-end slope from the luminosity distributions of
the sample using a maximum likelihood estimate.
Assuming a single-power law luminosity function:
\begin{equation}
\Psi(M_{1450}) = \Psi^* 10^{-0.4[M_{1450}+26](\beta+1)},
\end{equation}
we find that for the $\Lambda$-model,
$\Psi^* = (3.3^{+3.8}_{-1.6}) \times 10^{-9} \rm Mpc^{-3}$.
The best-fit bright-end slope is $\beta = -3.2$,
with a 68\% confidence range of [--2.5, --4.0] and
a 95\% confidence range of [--2.2, --4.2].
This slope is steeper than that measured in Paper II,
where the best-fit value was --2.3 with a 68\% range of [--1.6, --3.1].
The reason for this 1-$\sigma$ change is that all three new
quasars included in the sample have $z_{AB} \sim 20$, close
to the detection limit. 
Their inclusion, when correcting for the lower completeness at
the faint end, drives the best-fit luminosity function to steeper slopes.
For example, the average selection probability of quasars with
the redshift and luminosity of SDSS J1623+3112 is of
the order 10\%; the strong Ly$\alpha$ emission boosts the $z$-band
magnitude and makes the $z-J$ color bluer, making the selection of
this object easier than an average $z=6.2$ quasar.
The best-fit slope omitting SDSS J1623+3112 is --2.8.
The slopes derived here and in Paper II differ only at the 1-$\sigma$
level.  The errors are large because of the small number of objects in
the sample, especially at the faint end.  
This underlines the need for a large sample to put strong constraints on the quasar
luminosity function at the highest redshift.

The slope of the quasar luminosity function has
important implications.
The total quantity of ionizing photons emitted by the high-redshift quasar
population is determined by the quasar luminosity function.
Fan et al. (2001) found that $z\sim 6$ quasars are not likely
to be the sources responsible for reionizing the universe, or
keeping the universe ionized at high redshift, assuming 
a slope of $\beta > -3.5$.  Interestingly, we are approaching this
limit with the current analysis.  

In a flux-limited sample, 
the lensing probability of the brightest observed objects is boosted by
magnification bias (e.g. Turner et al. 1984).
The theoretical prediction of the
fraction of strongly-lensed quasars at high redshift 
could be be of order unity
for a sufficiently steep luminosity function (e.g. Wyithe \& Loeb 2002a,
Comerford, Haiman \& Schaye 2002). 
Fan et al. (2003) and
Richards et al. (2004) used the lack of multiply-imaged quasars
among the SDSS $z\sim 6$ quasar sample to constrain the shape of the
quasar luminosity function to be $\beta > -4.6$ at the 3-$\sigma$ level.
The lensing fraction increases by a factor of $\sim 5$ by assuming
a slope of $\beta = -3.3$ rather than $\beta = -2.2$ (Wyithe \& Loeb 2002b).
The quasar emissivity and lensing fraction also depends strongly on
the faint-end quasar slope, which is currently completely unknown.

In this paper, we present the first quasar discovered in the
faint quasar survey using co-added catalogs 
from the SDSS Southern Survey region.
In Fall 2003, we also started to use co-added {\em images} of the SDSS Southern Survey 
from $\sim 8$ SDSS observations to select candidates.
This overcomes the limitation of the co-added catalog which requires the
object to be detected in a single exposure.
We have generated a preliminary photometric catalog using the
SExtractor software; the integration of the co-added imaging into
the SDSS photometric pipeline is currently underway. 
Using the catalogs generated by
co-added imaging, we were able to recover  SDSS J000552.34--000655.8.
The remaining candidate identification is still in progress and will
be reported in the future. 
In the next few years, 
the faint quasars selected from the Southern Survey will be combined 
with the bright quasars in the Main Survey region to study 
the evolution of quasar population at $z\sim 6$ down to much  fainter
luminosities to probe the evolution of faint quasars.

Funding for the creation and distribution of the SDSS Archive has been provided by the Alfred P. Sloan Foundation, the Participating Institutions, the National Aeronautics and Space Administration, the National Science Foundation, the U.S. Department of Energy, the Japanese Monbukagakusho, and the Max Planck Society. The SDSS Web site is http://www.sdss.org/.
The SDSS is managed by the Astrophysical Research Consortium (ARC) for the Participating Institutions. The Participating Institutions are The University of Chicago, Fermilab, the Institute for Advanced Study, the Japan Participation Group, The Johns Hopkins University, Los Alamos National Laboratory, the Max-Planck-Institute for Astronomy (MPIA), the Max-Planck-Institute for Astrophysics (MPA), New Mexico State University, University of Pittsburgh, Princeton University, the United States Naval Observatory, and the University of Washington.
We thank the staffs at Apache Point Observatory, the MMT, the Bok
Telescope, Kitt Peak, the Hobby-Eberly Telescope, and Magellan for their
expert help.
We acknowledge support from NSF grant AST 03-07384, a Sloan Research Fellowship
and the University of Arizona (X.F.), NSF grants
AST 00-71091 and AST 03-07409 (M.A.S.) and NSF grants AST 99-00703 and 
AST 03-07582 (D. P. S.).

\begin{deluxetable}{cccccc}
\tablenum{1}
\tablecolumns{6}
\tablecaption{Photometric Properties of Five New $z>5.7$ Quasars}
\tablehead
{
object & redshift & $i$ & $z$ & $J$ & SDSS run
}
\startdata
J000239.39$+$255034.8 & 5.80 $\pm$ 0.02 & 21.47 $\pm$ 0.11 & 18.99 $\pm$ 0.05 & $>16.5$  & 4152 \\
J000552.34$-$000655.8 & 5.85 $\pm$ 0.02 & 23.40 $\pm$ 0.34 & 20.54 $\pm$ 0.10 & 19.87 $\pm$ 0.10 & multiple \\
J141111.29$+$121737.4 & 5.93 $\pm$ 0.02 & 23.43 $\pm$ 0.37 & 19.63 $\pm$ 0.07 & 18.95 $\pm$ 0.05 & 3836 \\
                    &                 & 22.85 $\pm$ 0.30 & 19.65 $\pm$ 0.08 &                  & 3996 \\
J160254.18$+$422822.9 & 6.07 $\pm$ 0.02 & 22.78 $\pm$ 0.38 & 19.89 $\pm$ 0.10 & 18.46 $\pm$ 0.05 & 3705 \\
J162331.81$+$311200.5 & 6.22 $\pm$ 0.02 & 24.52 $\pm$ 0.62 & 20.09 $\pm$ 0.10 & 19.15 $\pm$ 0.10 & 3918 
\enddata
\tablenotetext{}{The SDSS photometry ($i,z$) is
reported in terms of {\em asinh magnitudes} on the AB system.
The asinh magnitude system is defined by Lupton, Gunn \& Szalay (1999);
it becomes a linear scale in flux when the absolute value of the
signal-to-noise ratio is less than about 5. In this
system, zero flux corresponds to 24.4 and 22.8,
in $i$, and $z$, respectively; larger magnitudes refer to negative flux values.
The $J$ magnitude is on a Vega-based system.}
\end{deluxetable}

\begin{deluxetable}{ccccccc}
\tablenum{2}
\tablecolumns{7}
\tablecaption{Continuum Properties of new $z>6$ Quasars}
\tablehead
{
object & redshift & $AB_{1280}$ & $AB_{1450}$ & $M_{1450}$ & $M_{1450}$ & $E(B-V)$ \\
  &  &   &  & ($\Omega$-model) & ($\Lambda$-model) &  (Galactic)
}
\startdata
J000239.39$+$255034.8 & 5.80 & 19.09 & 19.02 & --27.40 & --27.66 & 0.037 \\
J000552.34$-$000655.8 & 5.83 & 20.30 & 20.23 & --26.21 & --26.46 & 0.033 \\
J141111.29$+$121737.4 & 5.93 & 20.04 & 19.97 & --26.49 & --26.75 & 0.025 \\
J160254.18$+$422822.9 & 6.07 & 19.93 & 19.86 & --26.63 & --26.82 & 0.014 \\
J162331.81$+$311200.5 & 6.22 & 20.20 & 20.13 & --26.40 & --26.67 & 0.022
\enddata
\end{deluxetable}

\begin{figure}[hbt]
\epsscale{0.70}
\plotone{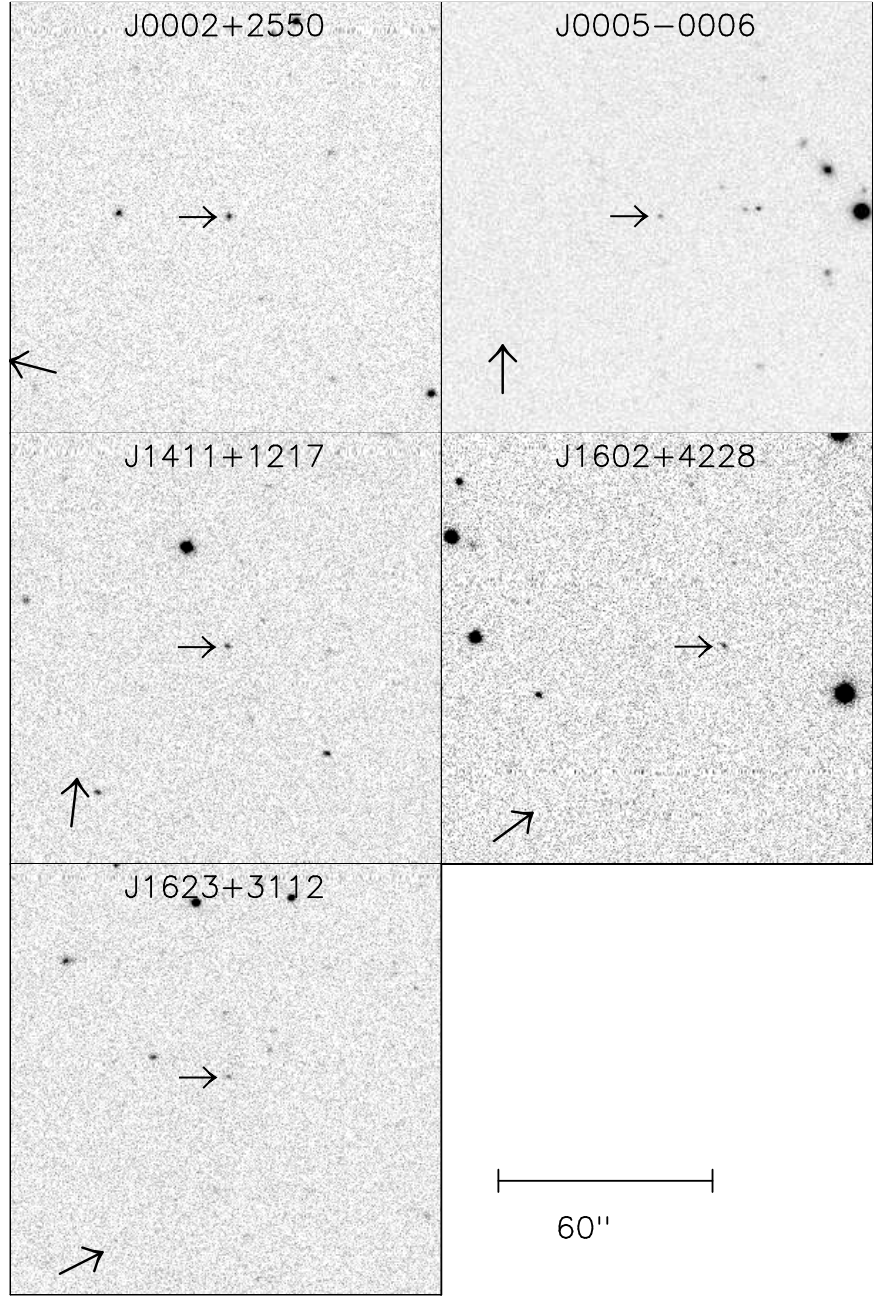}
\caption{SDSS $z$-band images of the five new $z>5.7$ quasars.
Each side of the finding chart is 160$''$. 
The arrow at the lower left indicates the direction of North on
the finding chart;
East is 90$^{\circ}$ counterclockwise from North.
}
\end{figure}

\begin{figure}[hbt]
\epsscale{0.70}
\plotone{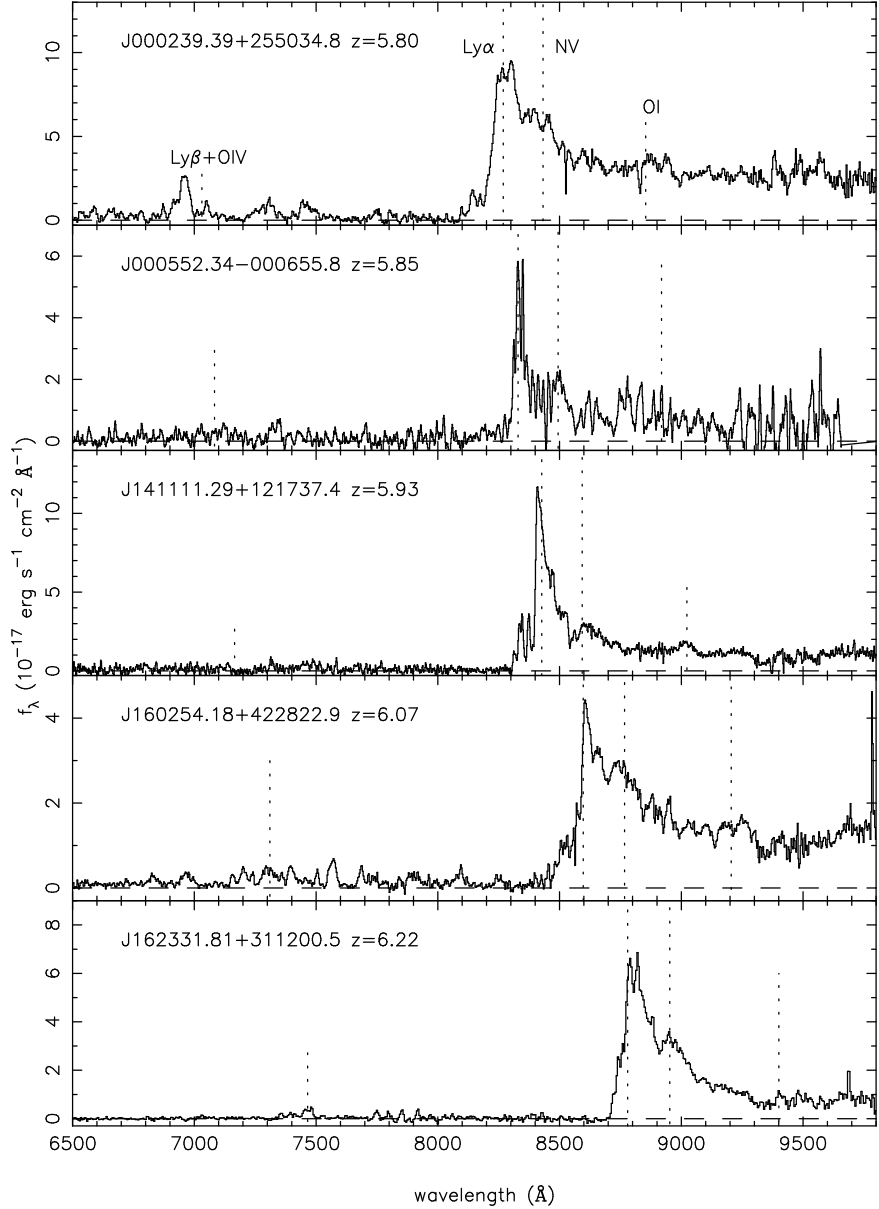}
\caption{Spectra of the five new quasars at $z>5.7$.
The spectrum of SDSS J0002+2550 is a 60-min exposure taken with the MMT-Red Channel;
the spectrum of SDSS J0005-0006 is a 40-min exposure taken with Magellan II and B\&C spectrograph;
the spectrum of SDSS J1411+1217 is a 120-min exposure taken with KPNO-4m and MARS; the spectrum of SDSS J1602+4228 is a 120-min exposure taken with the MMT-Red Channel; and the spectrum of SDSS J1623+3112 is a co-addition of two 120-min 
exposures using MARS and the Red Channel.
The fluxes are scaled to reproduce the $z$-band magnitude as measured
by the SDSS.
All spectra are binned to a dispersion of 5\AA\ per pixel; the spectral resolutions
are between 500 and 1000, depending on the spectrograph used.
}
\end{figure}

\begin{figure}[hbt]
\epsscale{0.80}
\plotone{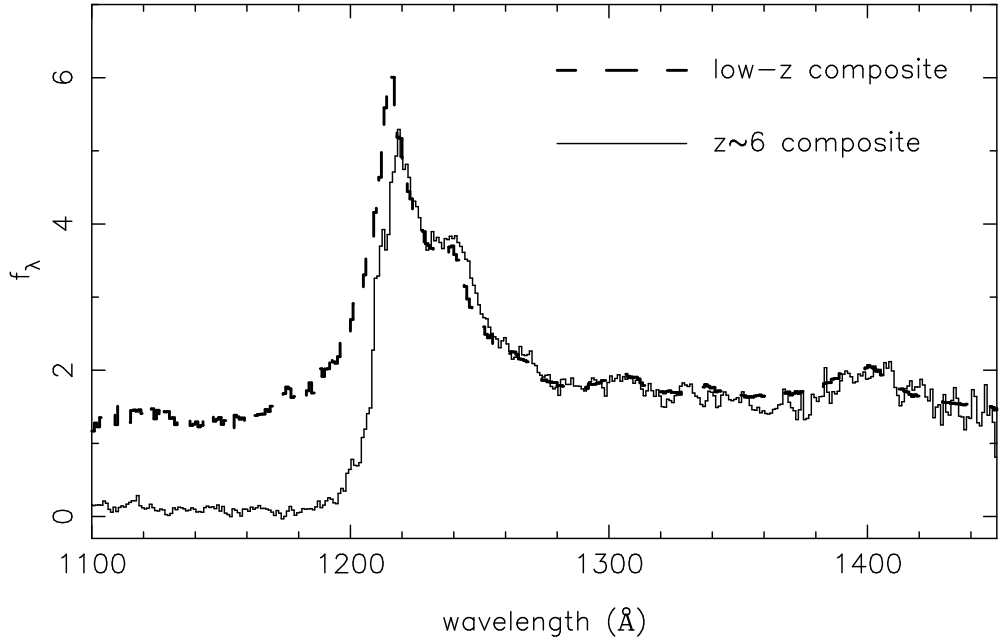}
\caption{
The composite spectrum of eleven $z\sim 6$ quasars (solid line). 
The spectrum of SDSS J0005-0002 was not included because of its low S/N.
The spectrum of each quasar is redshifted, scaled according to its
$m_{1450}$ magnitude, and  averaged with equal weighting.
For comparison, we also plot the low-redshift quasar spectral composite from
Vanden Berk et al. (2001). The effective redshift in the 1000 -- 1500\AA\ range
in the  Vanden Berk et al. composite is about 2.
The quasar intrinsic spectrum redward of Ly$\alpha$ emission shows
no detectable evolution up to $z\sim 6$, in terms of both the continuum
shape and emission line strengths. On the blue side of Ly$\alpha$
emission, the strong IGM absorption at $z\sim 6$ removes most of the quasar
flux. 
}
\end{figure}

\begin{figure}[hbt]
\epsscale{0.80}
\plotone{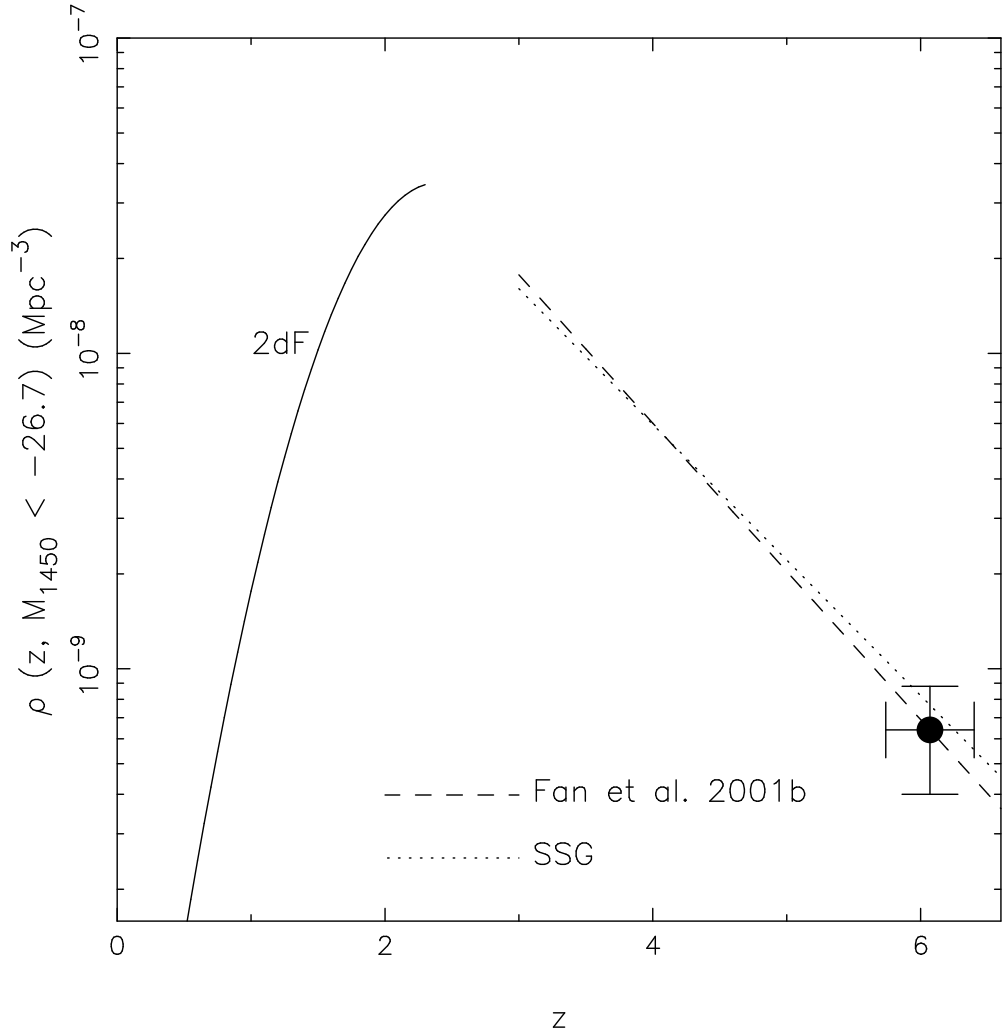}
\caption{
The evolution of the quasar comoving spatial density at
$M_{1450} < -26.7$ in the $\Lambda$-model.
The filled circle represents the result from this survey.
The error-bar in redshift indicates the redshift range
covered by the $i$-dropout survey.
The dashed and dotted lines are the best-fit models
from Fan et al.~(2001b) and Schmidt et al.~(1995, SSG),
respectively.
The solid line is the best-fit model from the 2dF survey
(Croom et al. 2004) at $z<2.3$.
}
\end{figure}
\newpage

\end{document}